\documentclass[aps]{revtex4}
\usepackage{graphicx}

\def\bea{\begin{eqnarray}}
\def\eea{\end{eqnarray}}

\def\sqr#1#2{{\vcenter{\vbox{\hrule height.#2pt
      \hbox{\vrule width.#2pt height#1pt \kern#1pt
         \vrule width.#2pt}
      \hrule height.#2pt}}}}

\def\figloc#1#2 {
\begin{figure}\begin{center}
    \includegraphics[width=80mm]{fig#1.ps}
    \caption{ #2}
    \end{center}\end{figure}
}

\begin{document}
\title{`` Prehawking" radiation  }

\author{W. G. Unruh}
\affiliation{ CIfAR Cosmology and Gravity Program\\
Dept. of Physics\\
University of B. C.\\
Vancouver, Canada V6T 1Z1\\
~
email: unruh@physics.ubc.ca}

~

~

\begin{abstract}
	Using the 2-D quantum energy momentum tensor expectation value near a
	black hole, the value near a collapsing shell which stops collapsing
	just outside the putative horizon is calculated and shown not to have
	anevidence of preHawking radiation.
\end{abstract}

\maketitle

Recently the issue of what has been called "prehawking"
radiation\cite{barcelo} has come to
be of interest in a number of papers\cite{mannetal}, especially whether or not the quantum
emission of radiation from a collapse which never forms a horizon could be
sufficient to explain that lack of a horizon. This seems to have begun by a
series of papers by Barcelo et al \cite{barcelo} where they raise the
possibility of Hawking radiation being produced in a collapse in which the
horizon never forms. 

Let us use the model in Chen {\it et al}\cite{chen} in which one has the collapse
of a shell of matter with the metric
\bea
ds^2 =\left\{ \begin{array} {lr} du^2 +2dudr -r^2 d\Omega^2 & r<R(u)\\
                     (1-{2m(u)\over r})U'(u)^2 du^2 +2 U'(u) dudr -r^2
	     d\Omega^2 & r>R(u) 
\end{array}
	     \right\}
\eea
Where we demand that the metric be continuous at $r=R(u)$ the path of the
shell. $U=U(u)$ is the usual Eddington-Finkelstein null Schwartzschild
parameter outside the shell. The continuity of the metric across the surface
$r=R(u)$ gives
\bea
R'(u)=- {U'(u)^2\left(1-{2m(u)\over R(u)}\right)-1\over 2(U'(u)-1)}
\eea

In the spirit of Barcelo {\it et al}\cite{barcelo} let us take m(u) to be a constant, and
the the shell to be a null shell so that $R(u)=2m+\epsilon-{u\over 2}$  
until $R(u)$ is within $\epsilon$ of $2m$ (ie, at u=0).i $\epsilon$ is assumed
to be very small ($<<2m)$). At that point we stop the shell so
$R(u)=2m+\epsilon$ thereafter. As Barcelo et al indicated, th Bogoliubov coefficients indicate that this metric will radiate Hawking
radiation until u=0. from about u=-m to u=0. During this time,
\bea U'={1\over(1-{2m\over (2m+\epsilon-u/2)})}= {2m+\epsilon -u/2\over \epsilon
-u/2}\approx {2m\over \epsilon -u/2}
\eea
or 
\bea
U(u)\approx  4m\ln\left({ \epsilon-u/2\over m}\right)
\eea

To get an estimate of the Energy Momentum tensor expectation value, we use the
two dimensional calculation of Davies, Fulling and Unruh (DFU) \cite {DFU} of the energy
momentum tensor of a massless scalar field in 2 spacetime dimension. This
calculation is easy, because the massless scalar field is conformally
invariant in 2 dimensions, and a null left going wave remains a null left
going wave no matter how wild the spacetime becomes. There is no
back-scattering by the field. Writing the metric in the form
\bea
ds^2= e^{2\xi(u,v)} dudv
\eea
which is always possible in two dimensions, and  assuming that for large
negative $u,v$ that the modes which go as $e^{i\omega u}$ or $e^{i\omega v} $
for $\omega>0$ are the modes associated with the annihilation operators of the
incoming vacuum state, then
\bea
T_{uu}= {1\over 48\pi}\left[2\xi_{,uu}-(\xi_{,u})^2\right]\\
T_{vv} = {1\over 48\pi}\left[2\xi_{,vv}-(\xi_{,v})^2\right]\\
T_{uv}= {1\over 48\pi}{\cal R}g_{uv} =-{1\over 24\pi} \xi_{,uv} 
\eea
where ${\cal R}$ is the scalar curvature. This energy momentum tensor is
conserved.
The off diagonal term is the so called conformal anomaly. Choosing the above
2-D metric 
\bea
ds^2=\left\{\begin{array}{lr}du^2+2dudr & r>R(u)
\\ U'(u)^2 (1-{2m\over r}) du^2 +2U'(dudr)&r<R(u)
\end{array} \right.
\eea
we rewrite this in the double null form

\bea
ds^2=\left\{\begin{array}{lr} U'(u) e^{(U(u) +V)/4m}\left({e^{r(U(u)-V)/4m}\over
	r(V-U(u)}\right) dudV
			&v(V)-u>R(u)\\
		v'(V) dudV& v(V)-u<2R
		\end{array}\right.
\eea
where $r(V-U(u))$ is defined by 
\bea
r(V,U(u))+2m\ln({r(V,U(u))-2m\over 2m})= {V-U(u)\over 2}
\eea

As shown in Chen {\it et al} paper, using the condition that the metric across
the shell is continuous, we have
\bea
R'(u)= -{{U'}^2(1-{2m\over R(u)}) -1\over 2(U'-1)} 
\eea
which for a null shell, where $R'(u)=-1/2$, gives
\bea
U'= {1\over 1-{2m\over R(u)}}
\eea
Since it is in the $u,V$ coordinates outside the shell that we have the
standard positive norm meaning positive frequency definition of the vacuum
state, the DFU comformal factor is 
\bea
e^{\xi}= {1-{2m\over r}\over 1-{2m\over R(u)}}
\eea
we see that along the shell, where $r=R(u)$, $\xi(u,0)=0$. Thus along the
shell, $T_{uu}=0$ There is no quantum emission from the shell into the region
outside the shell. 

For $v<0$, we have $V=v$. For $v>0$, the $v$ coordinates cross the shell after
it has stopped, and $v(V)= {\epsilon\over 2m}  V$ In both cases there is no flux of
radiation inwards beyond the shell although there is at $v=0$.

If $V>0,u<0 $ and $r>\epsilon$ this is precisely  the calculation in DFU {\it et al}
\cite{DFU}. Ie, for $u<0$, the energy momentum tensor is exactly what one would
expect for a black hole. There is a positive flux of energy  travelling out to infinity,
a negative energy flux heading toward the shell, and the conformal anomaly
acts as if it is the source of both of these fluxes. Ie, the outward flux dies
off for  $r$ less than about $3m$ and the negative ingoing flux dies of for
$r$ greater than around $3m$. These fluxes are all very very small (or order
$1/m^2$ in Planck units-- ie, about one photon of frequency the inverse
black-hole time, emitted per black hole timescale. They have a completely
negligible effect on the metric, except over very very long time scales (of
order $m^3$ in Planck units, or $10^53 $ ages of the current universe for a
solar mass black hole. They certainly have no effect on the behaviour of the
shell. 

For $u>0,v>0$, when the shell has stopped, but $u<u_0$ where $u_0>0$ is the value of $u$ where the null
line ingoing line from the turnaround reflects off $r=0$, the zero ingoing
($T_{vv}$) radiation produces zero outgoing radiation through the shell.
However, $T_{vv}$ is negative, and that negative ingoing flux produces a
negative outgoing $T_{uu}$ for $u>u_0$, equal to the negative ingoing $T_vv$
This is exactly the situation for the quantum energy momentum tensor in the
eternal Schwarzscild spacetime as was shown in DFU. In the usual
Schwartzschild coordinates, they showed that this gave an energy momentum
tensor with negative energy density near $r=2m$ (the negative energy density
of the Boulware vacuum). Ie, there is no prehawking radiation. The radiation
is all just the Hawking radiation for $u<u_0$ and the static Schwartzschild radition
for $u>u_0$. These are both tiny (total energy radiated going as $1/m^2$ in
Planck units.).

\figloc{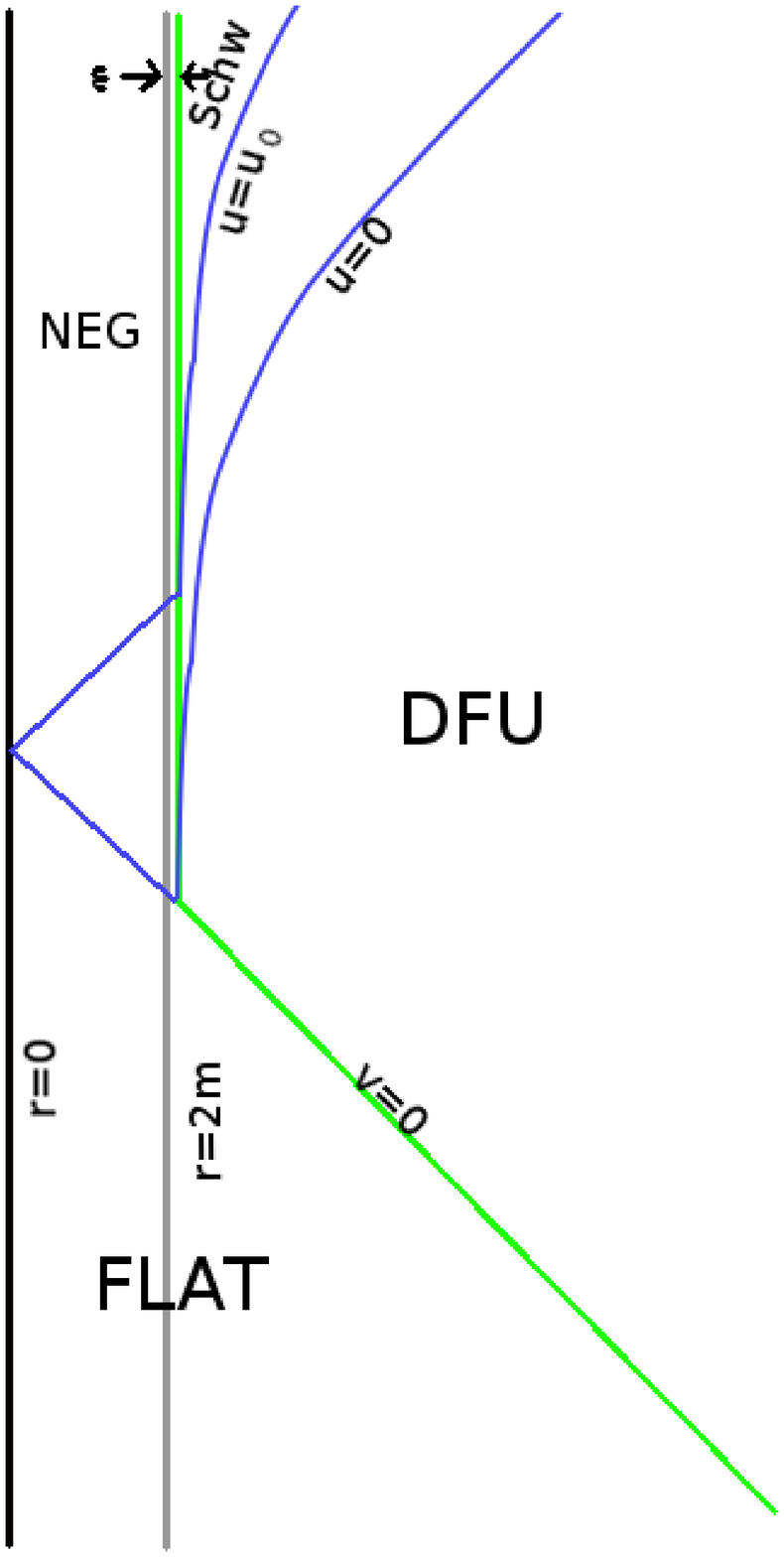}{This figure, in Eddington Finkelstein advanced
	coordinates (ingoing null rays are at 45 degrees), shows the various
	regions of the spacetime. The green lines designate the ingoing null
	matter which comes to a halt a distance $\epsilon$ from the "horizon"
	($r=2m$). The blue lines designate null surfaces. The first outgoing
	null line is the outgoing null line which connects to the point where
	the shell first comes to rest. The second outgoing blue line
	designates the outgoing curve which connects, via reflection from the
	origin, to the ingoing effect of that stopping point. The region
	labeled DFU is the region where the quantum energy momemtum tensor is
	what was calculated in the DFU paper, and is identical to what it
	would be if the collapse had actually produced a black hole. The area
	labeled Flat is flat spacetime, where the energy momentum tensor
	expectation value would be expected to be 0. The region labeled
	Schwarzschild is where the energy momentum tensor is what it would be
	in the Boulware vacuum of a black hole. Note that we would expect
	these regions to also be similar in the case of a 4-D system. The
	region labeled NULL is where the ingoing null flux impinging on the
	surface of the shell travels into the flat spacetimetime inside the
	shell, reflects from r=0 and goes out again producing a null outgoing
	flux outside the shell. }

However in order to make the shell behave in this way, one needs a horrendous
transverse stress within the shell. One can imagine what is needed by noting that
one has an infalling shell travelling at the velocity of light, and on a scale
far less than the Planck scale the shell must come to a stop. This would
require a transverse pressure far far larger than the energy in the
shell(drastically violate all of the energy limits).
In order to have the DFU wedge be a reasonable
size-- eg more than than say a second for a solar mass black hole, between the
infalling shell and the outgoing u=0 surface,  at a radius of a few 
Schwartzschild radii, $\epsilon$ would have to be of the order of $e^{-5000}$ of the
Planck time or distance.

Ie, if one tries to rely on the quantum emission to affect the trajectory of
the shell, even in this case where no horizon forms, the quantum emission is
simply far too small to produce any effect. One thus needs 
unphysical internal pressures, with no physical origin, to stop the collapse. 
If one stops the collapse well outside the surface $r=2m$ internal pressures
can clearly do so. The collapse of matter to form the earth was stopped well
outside the horizon by internal pressures. To stop it within some tiny,
sub-Planck distance from the horizon is however another matter entirely.

\end{document}